\documentclass{article}

\usepackage[utf8]{inputenc}

\usepackage{microtype, spconf}

\usepackage[numbers,square]{natbib}

\usepackage{booktabs, graphicx}
\usepackage{color}
\usepackage{url}

\title{Enhancing Situational Awareness in Wearable Audio Devices Using a Lightweight Sound Event Localization and Detection System}
\name{Jun-Wei Yeow$\sthanks{This research is supported by the Ministry of Education, Singapore, under its Academic Research Fund Tier 2 (MOE-T2EP20221-0014) and Tier 1 Grant: RG11/23}$,
      Ee-Leng Tan,
      Santi Peksi,
      Zhen-Ting Ong, 
      Woon-Seng Gan 
      }
\address{Smart Nation TRANS Lab, School of Electrical and Electronic Engineering, \\
Nanyang Technological University, Singapore \\
junwei004@e.ntu.edu.sg, \{etanel, speksi, ztong, ewsgan\}@ntu.edu.sg\\          
}

\begin{document}

\maketitle 

\begin{abstract}
    
Wearable audio devices with active noise control (ANC) enhance listening comfort but often at the expense of situational awareness. However, this auditory isolation may mask crucial environmental cues, posing significant safety risks. To address this, we propose an environmental intelligence framework that combines Acoustic Scene Classification (ASC) with Sound Event Localization and Detection (SELD). Our system first employs a lightweight ASC model to infer the current environment. The scene prediction then dynamically conditions a SELD network, tuning its sensitivity to detect and localize sounds that are most salient to the current context. On simulated headphone data, the proposed ASC-conditioned SELD system demonstrates improved spatial intelligence over a conventional baseline. This work represents a crucial step towards creating intelligent hearables that can deliver crucial environmental information, fostering a safer and more context-aware listening experience.
\end{abstract}

\section{Introduction}
\label{Section:Introduction}

In recent years, the proliferation of wearable audio devices, or hearables, has transformed the field of personal audio by offering users immersive listening experiences~\citep{gupta2022augmented}. A key technology driving this advancement is Active Noise Control (ANC), which helps to effectively attenuate background noise. However, this noise suppression can significantly diminish the situational awareness of the listener. By masking crucial environmental sounds, ANC can inadvertently place users at risk by rendering them oblivious to potential hazards, such as approaching vehicles.   

To address this safety concern, researchers have begun to explore intelligent hearable technologies to manage the trade-off between auditory immersion and environmental awareness. For example,~\citet{veluri2023semantic} introduced `Semantic Hearing', enabling the selective extraction of target sounds using a pair of headphones. While innovative, this approach requires the users to manually define sounds of interest, limiting its adaptability in dynamic environments. Similarly, systems like the Pedestrian Audio Wearable System (PAWS) have been developed to detect specific threats, such as car horns~\cite{de2018paws}. However, such specialized systems often lack the flexibility needed to generalize to a broader range of important environmental sounds and may require cumbersome, non-standard hardware. 

\begin{figure*}[t]
\begin{center}
\includegraphics[width=\textwidth]{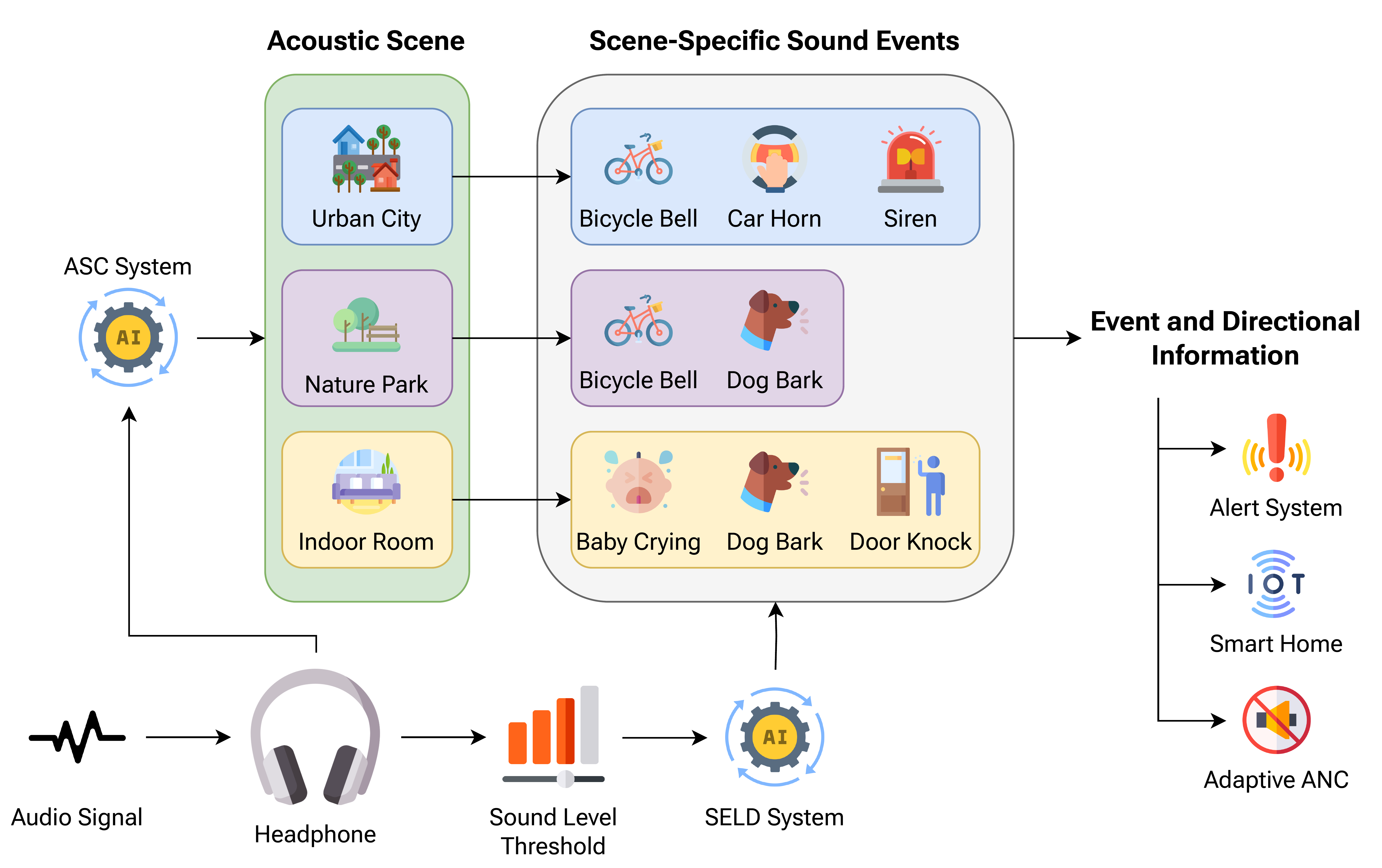}
\caption{Overview of the proposed ASC-conditioned SELD framework. The ASC front-end processes single-channel audio to predict the acoustic scene. These scene predictions are then used to adapt the SELD back-end system, which processes multi-channel audio to output the activity and direction of relevant sound events.}
\label{fg:Overview}
\end{center}
\end{figure*}

In this paper, we propose a novel environmental intelligence framework for hearables that integrates Acoustic Scene Classification (ASC) and Sound Event Localization and Detection (SELD). A standalone SELD system, while identifying and locating sound sources, fundamentally lacks contextual awareness. It applies a uniform detection strategy across all environments, which can reduce its effectiveness. Our approach overcomes this limitation by first using a lightweight ASC model to identify the user's current acoustic environment (e.g., \textit{Indoor}, \textit{Nature}, \textit{Urban}). This contextual information is then used to condition the SELD network, dynamically tuning it to prioritize events that are most salient to the identified scene. For instance, in an \textit{Urban} environment, the system would prioritize the detection of \textit{car horns} rather than \textit{crying} sounds. This scene-conditioning helps to improve the reliability and accuracy of detecting critical sounds, leading to more robust and practical spatial intelligence. Figure~\ref{fg:Overview} depicts the overall workings of our proposed ASC-SELD framework.

This context-aware approach effectively enables the hearable device to automatically and seamlessly provide spatial audio intelligence without requiring manual user input. Furthermore, the system outputs essential metadata, including acoustic scene, event activity, and Direction-of-Arrival (DOA) information. This can be leveraged by downstream applications to alert the user, enable selective hear-through, or integrate with other smart devices in the Internet of Things (IoT).

\section{Methods}
\label{section:Methods}

\subsection{Dataset Curation}
\label{subsect:Dataset}

\begin{figure*}[t]
\begin{center}
\includegraphics[width=\textwidth]{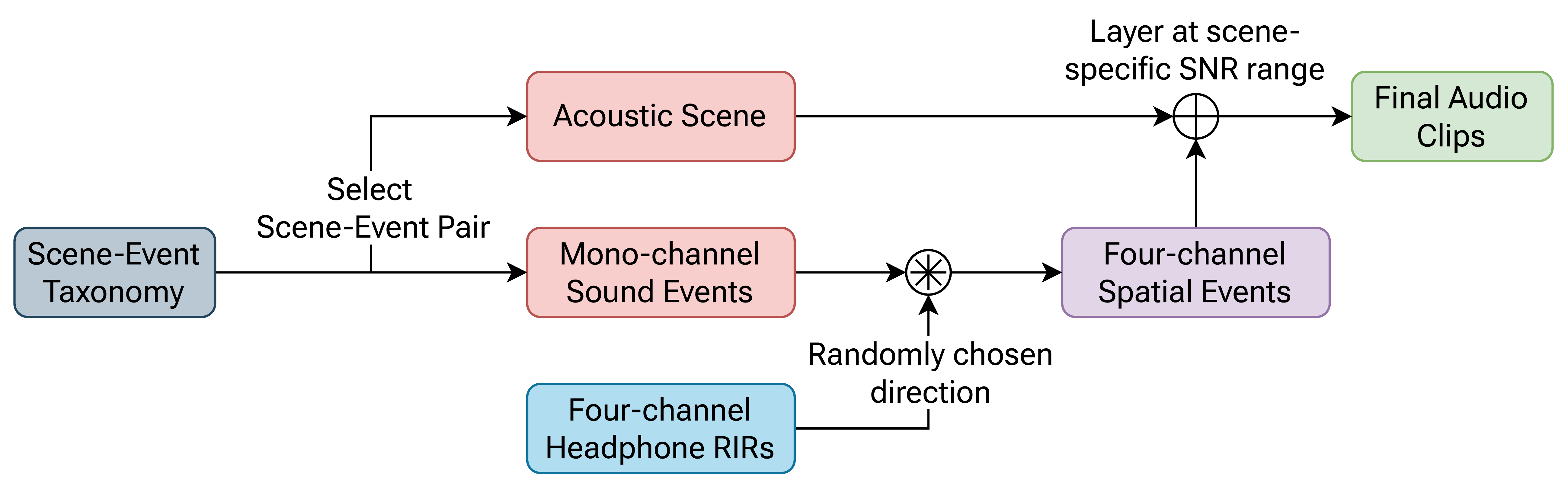}
    \caption{Block diagram of the synthetic data generation pipeline. Background scenes and anechoic sound events are combined and spatialized using headphone-specific RIRs to create labeled multi-channel audio clips for training and evaluation.}
\label{fg:dataset_gen}
\end{center}
\end{figure*}

The collection and annotation of real-world, multi-channel audio for SELD is costly and labor-intensive. To overcome this, we instead use a synthetic data generation pipeline to create a comprehensive and diverse dataset. Figure~\ref{fg:dataset_gen} illustrates the overall data generation process employed in this study.

The data synthesis process begins by selecting a background audio segment from established acoustic scene datasets, namely the TAU Urban Acoustic Scenes 2020 Mobile dataset~\citep{heittola2020acoustic} and the CochlScene dataset~\citep{jeong2022cochlscene}. These background recordings are chosen to represent distinct acoustic environments broadly categorized as \textit{Indoor}, \textit{Nature}, and \textit{Urban}. Concurrently, individual sound event clips corresponding to the classes of \textit{Bicycle, Car Horn, Crying, Dog, Door Knock,} and \textit{Siren} are randomly selected from large-scale, publicly available audio collections such as FSD50K~\citep{fonseca2021fsd50k}, NIGENS~\citep{trowitzsch2019nigens}, and UrbanSound~\citep{salamon2014dataset_urbansound}. We choose these sound event classes for their prominence and relevance in everyday environments, as well as their potential impact on situational awareness for hearable device users. Table~\ref{table:DatasetSynthesis} provides a summary of the acoustic scenes and corresponding sound events selected for inclusion. 

Following the selection process, each chosen sound event is convolved with a multi-channel Room Impulse Response (RIR) from a specialized dataset captured explicitly for wearable audio devices~\cite{corey2019acoustic}. These RIRs were recorded at uniform azimuth intervals of $15^\circ$, covering a complete $360^\circ$ horizontal plane, resulting in 24 discrete DOAs. These spatialized sound events are then mixed into the selected acoustic scene background at predefined, scene-dependent Signal-to-Noise Ratio (SNR) levels. The final dataset comprises a total of 10,800 training and 1,800 testing clips, each with a duration of one second at a sampling rate of $24\,\mathrm{kHz}$, simulating the microphone configuration on a standard pair of headphones. 

\begin{table}[t]
\caption{Description of the acoustic scene classes and the corresponding sound events of interest in each scene.}
\begin{center}
\begin{tabular}{ccc}
\toprule
Scene & Sound Events & SNR Range \\
\midrule
Indoor &  Crying, Dog, Door Knock   & [5, 20] \\
Nature &  Bicycle, Dog         & [0, 15] \\
Urban  &  Bicycle, Car Horn, Siren     & [-10, 5] \\
\bottomrule
\end{tabular}
\label{table:DatasetSynthesis}
\end{center}
\end{table}

\subsection{Deep Learning Models}
\label{subsect:DL_Models}

A primary consideration for developing our framework for hearables is the limited computational resources of these devices. Therefore, the underlying deep learning models must be lightweight and efficient. The field of computational audio analysis has seen significant progress in developing low-complexity models, driven largely by challenges such as the Detection and Classification of Acoustic Scenes and Events (DCASE). Building on these developments, our ASC front-end uses the highly efficient and effective model proposed by~\citet{Tan2025}, which recently achieved $2^\mathrm{nd}$ place in the DCASE 2025 low-complexity ASC task.

\begin{figure}[t]
\begin{center}
\includegraphics[width=0.75\columnwidth]{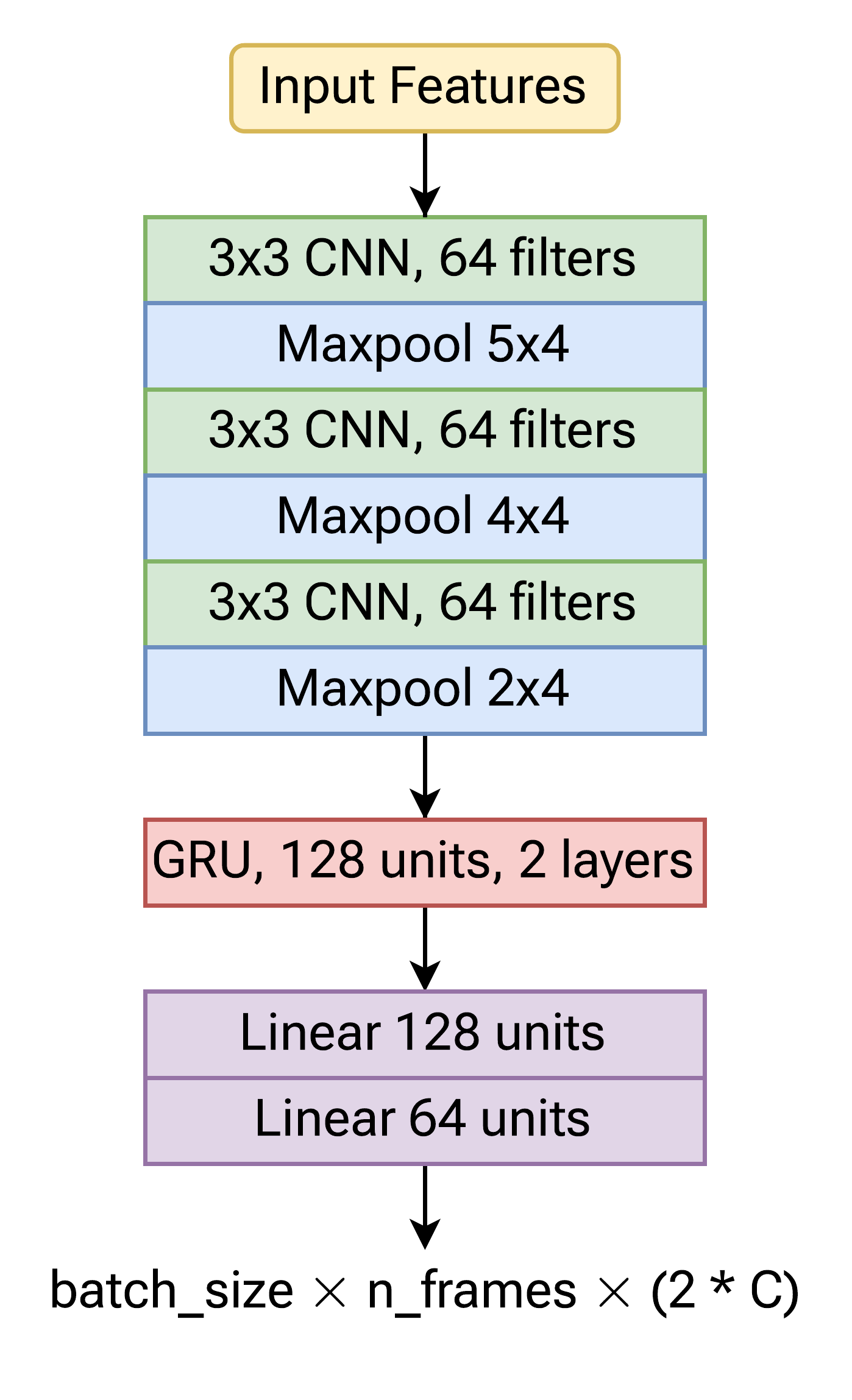}
    \caption{Block diagram of the SELDNet architecture adopted for this work. Maxpool operations are shown in terms of time \(\times\) frequency bins.}
\label{fg:seldnet_blockdiag}
\end{center}
\end{figure}

In contrast, resource-efficient SELD systems tailored explicitly for low-power wearable devices are relatively underexplored. A recent study by~\citet{yeow2024real} revealed that the choice of input features often imposes a greater computational burden than the model architecture itself. Guided by these insights, our SELD back-end is based on the well-established SELDNet architecture proposed by~\citet{adavanne2018sound}. SELDNet comprises a convolutional recurrent neural network (CRNN) structure, as illustrated in Figure~\ref{fg:seldnet_blockdiag}. 

More importantly, we leverage the Spatial cue-Augmented Log-SpectrogrAm Lite (SALSA-Lite) feature set~\citep{nguyen2022salsalite}, which combines log-power linear spectrograms with normalized inter-channel phase differences. As demonstrated by~\citet{yeow2024real}, this provides an optimal trade-off between computational overhead and SELD performance. The combined SELDNet and SALSA-Lite system results in a lightweight yet robust SELD back-end suitable for online inference on edge hardware. Table~\ref{table:Model_Complexities} summarizes the computational complexity and parameter counts of the utilized models.

\begin{table}[t]
\caption{Complexity and size of the deep learning models. Multiply-Accumulate operations (MACs) are estimated for a 1-second input audio clip.}
\begin{center}
\begin{tabular}{clcc}
\toprule
Model &  & MACs (M) & Params (K) \\
\midrule
ASC & \cite{Tan2025} & 10.9 & 116 \\
SELD & \cite{adavanne2018sound} & 91.4 & 285 \\
\bottomrule
\end{tabular}
\label{table:Model_Complexities}
\end{center}
\end{table}

\subsection{Scene-Conditioned Thresholding}
\label{subsect:Scene-Conditioning}

The SELDNet model employs the Activity-Coupled Cartesian DOA (ACCDOA) representation introduced by~\citet{shimada2021accdoa}. At each time frame $t$, the network outputs a ($C\times2$)-dimensional vector, where $C$ is the number of classes:

\begin{equation}
    \label{eqn:ACCDOA}
    \mathbf{v}(t) = [x_1(t), \dots, x_C(t), y_1(t), \dots, y_C(t)]^\mathsf{T}.
\end{equation}

For each class $i$, the 2-D Cartesian vector $\mathbf{v}_i(t)=[x_i(t),y_i(t)]$ encodes direction and its Euclidean norm $r_i(t) = ||\mathbf{v}_i(t)|| \in [0, 1]$ indicates event activity. In conventional ACCDOA decoding, a single static threshold (typically set at $0.5$) is applied to $r_i(t)$ across all classes to binarize event activity. 

In contrast, our proposed system employs a more contextually-informed approach, utilizing a dynamic thresholding strategy conditioned on the predicted acoustic scene. In our scene-conditioned framework, we replace the global threshold with a lookup table of class- and scene-specific thresholds $\tau_{s,i}$. For a given acoustic scene $s$, a class $i$ is considered active only if $r_i(t) > \tau_{s,i}$. This allows the system to prioritize detection sensitivity for events contextually relevant to the current acoustic scene, while setting higher thresholds for events that are statistically unlikely within that environment.

By conditioning thresholds based on the acoustic environment, this approach improves precision on context-relevant events without completely excluding unexpected yet genuine sounds. One key advantage of this approach is its adaptability; new scenes, events, or user preferences can be accommodated by simply updating the threshold matrix, avoiding the need for complete network retraining. For this work, the threshold matrix values were determined through a grid search.

\subsection{Training and Evaluation}

The ASC front-end was trained following the procedure detailed in~\citep{Tan2025}. Input audio clips were converted into log-power Mel spectrograms using a Short-Time Fourier Transform (STFT) with a window size of 4096 samples and a hop length of 800 samples, while using 256 Mel bins. The model was trained for 150 epochs, utilizing the Adam optimizer with a batch size of 256.

The SELD system was trained independently, also employing the Adam optimizer. Training was performed for a total of 100 epochs with a batch size of 64. A two-phase learning rate schedule was employed, with a linear warm-up for the first 10 epochs to a peak learning rate of $1\times10^{-3}$, followed by a cosine annealing decay for the remaining 90 epochs. Following the original implementation by~\citet{nguyen2022salsalite}, SALSA-Lite features were extracted using a 512-point FFT with a Hann window of the same size and a hop size of 300 samples. The spatial aliasing frequency was set to $1\,\mathrm{kHz}$, corresponding to the head spacing of $18\,\mathrm{cm}$. 

Performance evaluation for the ASC module was based on the classification accuracy. To holistically evaluate the SELD system, we adopt the location-dependent F-score, a standard metric from the DCASE Challenges~\citep{politis2020overview}. Here, an event is considered a true positive only if both the event class is correctly predicted \emph{and} the estimated DOA is within a specified angular threshold of the ground truth. For this work, the angular threshold is set to $7.5^\circ$, which represents half the angular separation between adjacent RIRs in our dataset. We report the macro-averaged F-score across all event classes, denoted as $\mathrm{F}_{\leq 7.5^\circ}$.

\section{Results}

\subsection{Acoustic Scene Classification}

The ASC module was first evaluated independently to assess its suitability for deployment on wearable audio devices. Training and evaluation were carried out using our synthesized dataset, where only the first-channel from the four-channel synthetic recordings was used to simulate monophonic audio.

The ASC model achieved a classification accuracy of $91.7\%$ on the test set. This result indicates a robust capacity to accurately determine the acoustic context despite operating on brief audio segments. Notably, given that real-world acoustic scenes typically do not change instantaneously, performance can easily be enhanced by aggregating predictions over a longer temporal window (e.g., 10 seconds). Although such methods were not evaluated in this study, it is reasonable to anticipate that leveraging longer context windows would yield substantial gains in ASC accuracy~\citep{martin2022low}, thereby potentially providing even greater improvement for the downstream scene-conditioned SELD system.

\subsection{Sound Event Localization and Detection}

\begin{table}[t]
\caption{SELD performance under different contextual conditioning scenarios. The Oracle represents a perfect ASC system, providing an upper-bound performance.}
\begin{center}
\begin{tabular}{lcc}
\toprule
Experiment & $\mathrm{F}_{\leq 7.5^\circ}\;(\uparrow)$ & $\mathrm{ASC\;Acc.}\;(\uparrow)$ \\
\midrule
Baseline & 79.39 & - \\
Baseline + ASC & 80.78 & 91.7\% \\
Baseline + Oracle & 81.02 & 100\% \\
\bottomrule
\end{tabular}
\label{table:SELD_Results}
\end{center}
\end{table}

Table~\ref{table:SELD_Results} details the results of the back-end SELD system under different contextual conditioning scenarios. The baseline SELD model employed a global threshold of $0.5$ across all classes for event detection, representing the conventional approach devoid of contextual information. This baseline achieved a respectable $\mathrm{F}_{\leq 7.5^\circ}$ score of $79.39$, indicating reasonable performance across diverse acoustic environments.

\begin{figure}[t]
\begin{center}
\includegraphics[width=\columnwidth]{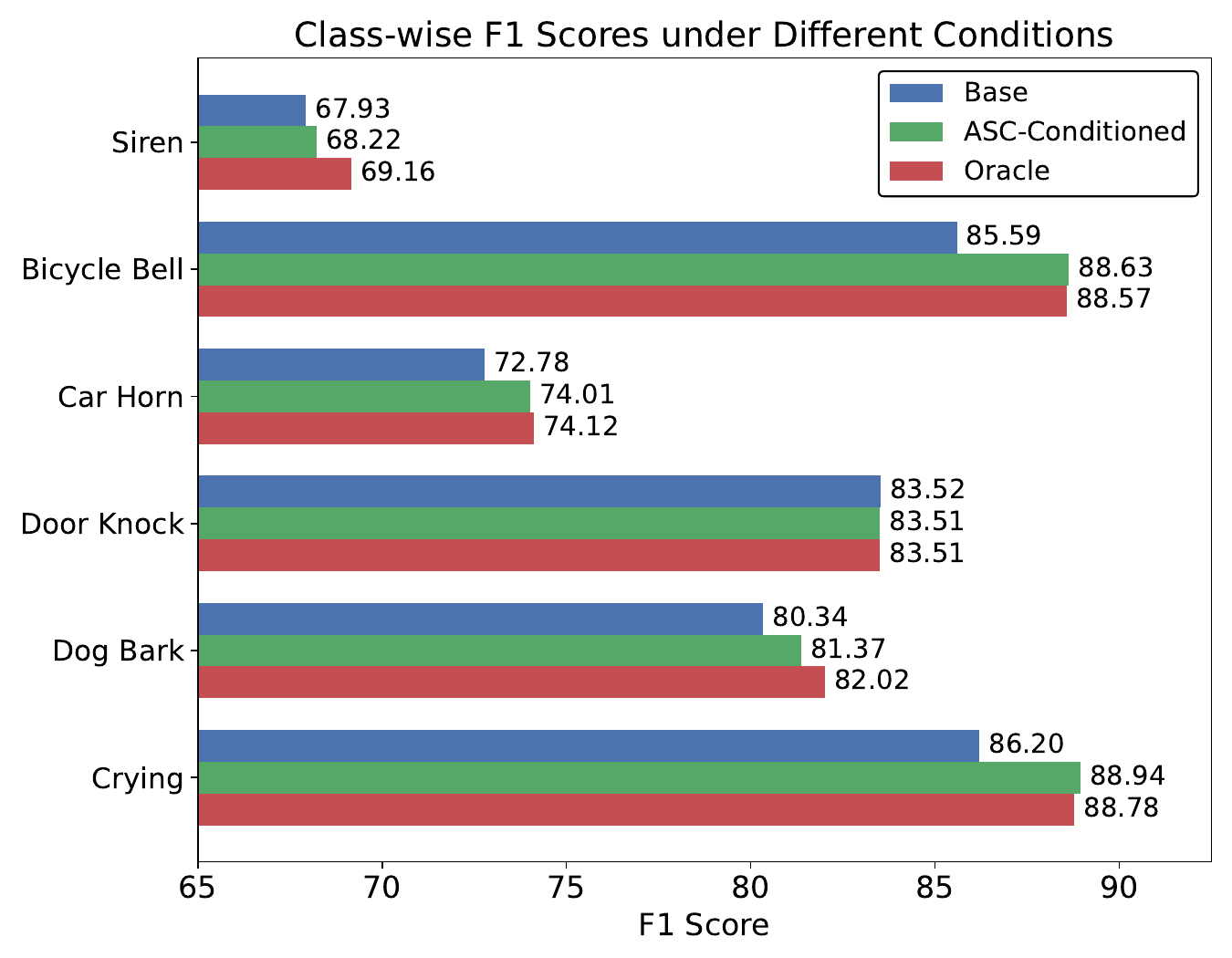}
    \caption{Class-wise breakdown of the F-score under different conditioning setups.}
\label{fg:fscore_breakdown}
\end{center}
\end{figure}

Subsequently, we evaluated our proposed method, which integrates the ASC predictions to dynamically adjust detection thresholds based on environmental context. With scene-conditioning, this proposed framework (Baseline + ASC) demonstrated improved performance, increasing the $\mathrm{F}_{\leq 7.5^\circ}$ score to $80.78$. Under an idealized scenario with perfect contextual knowledge (Baseline + Oracle), the ground truth scene labels are provided to the SELD system. This simulates a perfect ASC system and sets a practical upper-bound performance for our conditioning method. Under these conditions, the SELD system attained a $\mathrm{F}_{\leq 7.5^\circ}$ score of $81.02$. Figure~\ref{fg:fscore_breakdown} summarizes the class-wise $\mathrm{F}_{\leq 7.5^\circ}$ scores for each contextual conditioning scenario.

These performance gains across both experimental setups demonstrate the applicability and effectiveness of contextual intelligence to refine the SELD process. Moreover, the modest difference between the `Oracle' scenario and the ASC-conditioned setup suggests that imperfect scene information does not drastically degrade SELD performance. This can be attributed to our `soft conditioning' approach, where the ASC output is used to modulate detection thresholds, rather than as a `hard gate' that completely disables potential event detections.

\subsection{System Latency}

\begin{figure}[t]
\begin{center}
\includegraphics[width=\columnwidth]{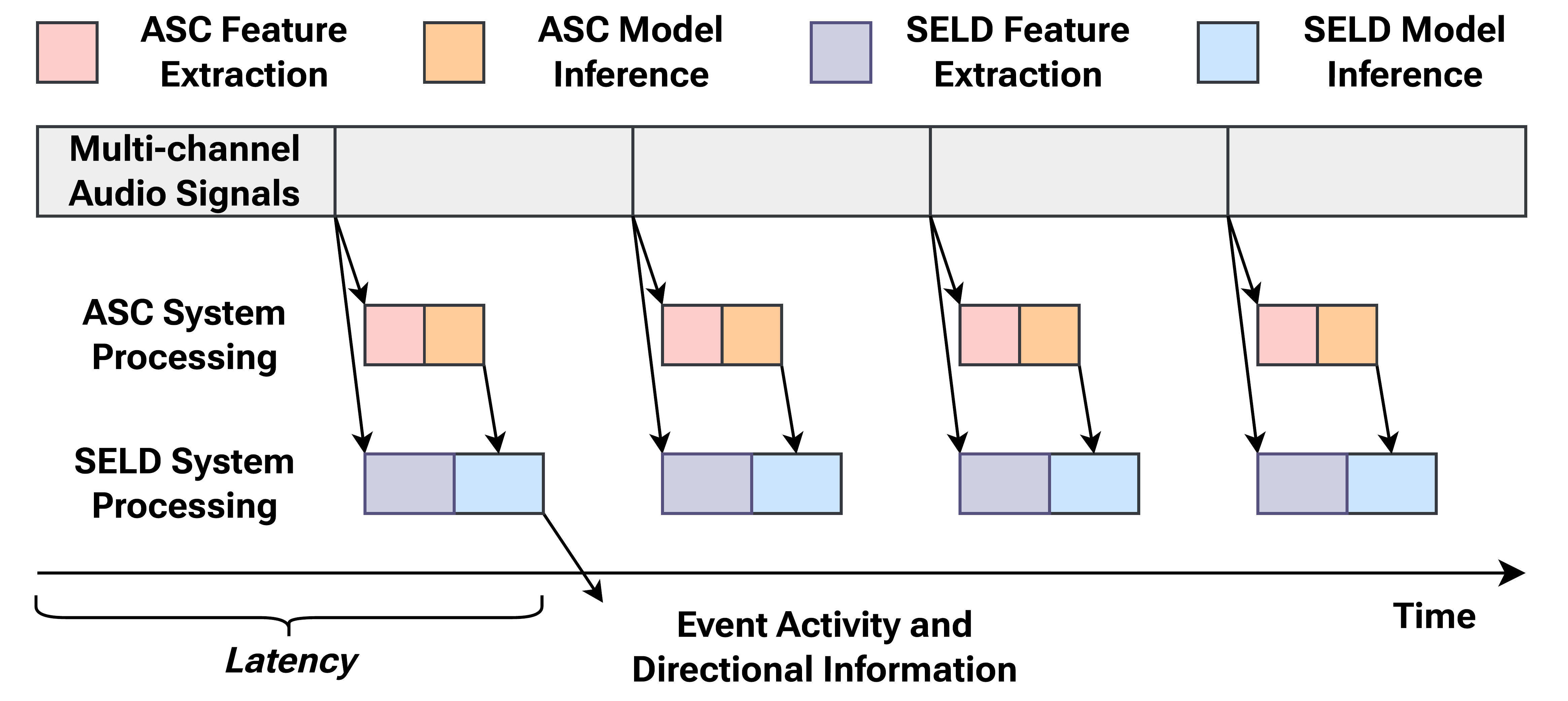}
    \caption{System latency details~\citep{veluri2023semantic}. Audio recording, ASC processing, and SELD processing all operate on separate threads.}
\label{fg:System_latency}
\end{center}
\end{figure}

\begin{table}[t]
  \centering
  \caption{Latency breakdown for each stage of the pipeline (all in milliseconds).}
  \label{tab:latency_breakdown}
  \begin{tabular}{l r}
\toprule

\multicolumn{2}{c}{\textbf{Acoustic Scene Classification}} \\
\midrule
Feature extraction        &  7.6 \\
Model inference           &  5.5 \\
\midrule
\textbf{ASC total}        & 13.1 \\
\midrule

& \\

\multicolumn{2}{c}{\textbf{Sound Event Localization \& Detection}} \\
\midrule
Feature extraction        & 18.5 \\
Model inference           & 19.5 \\
\midrule
\textbf{SELD total}       & 38.0 \\

\bottomrule
  \end{tabular}
\end{table}

For a system designed to enhance situational awareness, low latency is not only a performance metric but also a critical safety requirement. Therefore, we evaluate the processing latency of our integrated framework to determine its feasibility for real-time deployment on resource-constrained hardware. The system was benchmarked on a Raspberry Pi 4, a representative, commercially available edge device.  

As illustrated in Figure~\ref{fg:System_latency}, our framework is implemented using a multi-threaded architecture, adapted from~\citet{veluri2023semantic}. The audio acquisition, ASC processing, and SELD processing operate on separate, parallel threads. The latency breakdown for each stage of the inference pipeline, measured for a single one-second audio input averaged over 1,000 inputs, is presented in Table~\ref{tab:latency_breakdown}. 

From Table~\ref{tab:latency_breakdown}, the time taken for ASC processing (13.1 ms) is less than the time needed for SELD feature extraction (18.5 ms). This low latency processing opens up potential future avenues for more sophisticated integration of environmental information, rather than a soft conditioning mask approach presented in this work. Crucially, the end-to-end latency from the audio input to a potential SELD output is 38.0 ms. This low latency ensures that the listener will receive rapid and timely alerts about potential oncoming hazards, such as car horns~\citep{veluri2023semantic}. This further reinforces the applicability of our integrated framework for real-time, advanced spatial intelligence on resource-constrained hearing devices.

\section{Discussion}

The experimental results presented in this study demonstrate that integrating contextual information via an ASC front-end can enhance SELD performance. By adaptively modulating detection thresholds based on the inferred acoustic environment, the proposed framework improved SELD performance compared to a conventional static threshold baseline. The improvement in event detection reliability represents a meaningful step towards creating more intelligent and practical hearable devices. Moreover, the close performance between the ASC-conditioned and Oracle setup also illustrates the resilience of the proposed soft conditioning approach to occasional classification errors. This balance between contextual sensitivity and safety-critical reliability proves essential for real-world applications.

Nevertheless, we must acknowledge that the dataset used for this study is inherently synthetic, and not reflective of real-world conditions. Synthetic data, while useful for creating various experimental conditions, may not fully replicate the complexity of real-world acoustic conditions~\citep{roman2024spatial}. Real-world audio often contains non-stationary sound sources, a higher density of overlapping sound events, and complex reverberation patterns~\citep{shimada2023starss23}. This results in a `synthetic-to-real' domain shift problem that complicates real-world deployment~\citep{yeow2025enhancing}. Therefore, while our results provide a strong proof-of-concept for our spatial intelligence framework, validating the system under real-world conditions is a crucial next step for future research.

Finally, this work opens up promising avenues for future research, particularly in the optimization and personalization of the scene-event threshold matrix. While class-dependent conditioning improves performance~\citep{turpault2019sound}, these threshold values were manually set based on a grid search in our current implementation. Such manual threshold tuning may not scale efficiently to more diverse acoustic event sets. As such, an important direction for future research involves developing automated, data-driven optimization techniques to efficiently tune thresholds depending on the context.

\section{Summary}

In this paper, we introduced a lightweight and efficient framework for enhancing situational awareness in hearable devices by integrating ASC with SELD. Our approach uses a low-complexity ASC model to infer the user's acoustic environment. This contextual information is then used to dynamically adapt or condition a subsequent SELD network, allowing the system to focus on relevant environmental sounds. This novel scene-conditioning approach demonstrates improved detection performance, leading to more robust spatial intelligence. 

Experimental results on a synthesized dataset simulating headphone audio demonstrated that our ASC-conditioned SELD system outperforms a standard SELD baseline. The proposed framework represents a significant step towards creating safer, smarter, and more context-aware hearable devices. Future work will focus on acquiring real-world audio and physical deployment onto embedded hardware.

\section{Acknowledgment}

This research is supported by the Ministry of Education, Singapore, under its Academic Research Fund Tier 2 (MOE-T2EP20221-0014). 

\bibliographystyle{jaes}

\bibliography{refs}

\end{document}